\documentclass[12pt]{article}
\usepackage{enumerate}
\usepackage{amsmath}
\usepackage{amsfonts}
\usepackage{amssymb}
\newcommand{\bt}{\begin{theorem}}

\newcommand{\ds}{\displaystyle}
\newcommand{\et}{\end{theorem}}
\newcommand {\be}{\begin{equation}}
\newcommand {\ee}{\end{equation}}
\newcommand{\vsp}{\vskip 1em}
\newcommand{\hsp}{\hskip 2em}

\newcommand{\noi}{\noindent}
\newcommand{\wh}{\widehat}
\def \qed {\hfill \vrule height6pt width6pt depth0pt}

\title{Cooperative oligopoly games with boundedly rational firms}

\begin{document}
\date{}
\author{Paraskevas V. Lekeas{\footnote{SimplyHeuristics,
Chicago, IL 60631, USA; email: plekeas@simplyheuristics.com\vspace{0.2cm}}}\hsp Giorgos
Stamatopoulos{\footnote{Department of Economics, University of
Crete, 74100 Rethymno, Crete, Greece; email:
gstamato@uoc.gr\vspace{0.2cm}}}}

\maketitle

\begin{abstract}\noi We analyze cooperative Cournot games
with boundedly rational firms. Due to cognitive constraints,
the members of a coalition cannot accurately predict the
coalitional structure of the non-members. Thus, they
compute their value using simple heuristics.
In particular, they assign various non-equilibrium probability distributions
over the outsiders' set of partitions.
We construct the characteristic function of a coalition in such an environment and we analyze the core of the corresponding games. We
show that the core is non-empty provided the number of firms in the market is sufficiently large.
Moreover, we show that if two distributions over the set of partitions
are related via first-order dominance,
then the core of the game under the dominated distribution is a subset of the core
under the dominant distribution.

\end{abstract}

\vspace{0.2cm} \noi\hspace{0.83cm} {\small{\emph{Keywords}:
Cooperative game; externalities; Cournot market; core;
bounded rationality

\vspace{0.2cm} \hspace{0.33cm}\noi \emph{JEL Classification}: C71, L2}}

\section{Introduction}
Collusion among firms in oligopolistic markets is a wide-spread phenomenon
and constantly attracts the interest of economists. By colluding,
firms can restrict output and raise market prices, thus extracting a higher surplus from consumers.
From a methodological point of view, economists analyze collusion using either non-cooperative games
(when agreements among firms are non-enforceable by an outside entity) or
cooperative games (whenever the signing of enforceable agreements is possible). Under the latter approach, the focus usually is
on the core of an appropriately defined cooperative game. The core consists of all allocations of
total profits that cannot be blocked by any coalition of firms. Non-empty core means that cooperation among
all firms in the market is a priori feasible.

When a coalition of firms contemplate the blocking of an agreement,
they have to calculate their stand-alone payoff. In a market environment such a calculation is
not a trivial task, as the coalition's worth depends on how the non-members act. In particular, it depends on
the partition (coalition structure) that the outsiders will form. This calls for the formation of beliefs
about the outsiders' coalitional actions.

Different conjectures about the reaction of the outsiders lead to different coalitional worths and thus to different notions of core.
The $\alpha$ and $\beta$ cores (Aumann 1959) are based on the assumption of min-max behavior on behalf of the non-members;
the $\gamma$-core (Chander $\&$ Tulkens 1997) is based on the assumption that outsiders play individual best replies to the
deviant coalition; the $\delta$-core scenario (Hart $\&$ Kurz 1983) assumes that outsiders form a single coalition.
Various authors applied these core notions to the study of Cournot markets. Rajan (1989)
used the concept of $\gamma$-core and showed that it is non-empty
for a market with 4 firms. A more general result for any number of firms is provided by Chander (2010).
Currarini $\&$ Marini (2003) built a refinement of the $\gamma$-core by assuming that
the deviant coalition acts as a Stackelberg leader in the product market. Zhao (1999) showed that the $\alpha$ and $\beta$
cores of oligopolistic markets are non-empty.

The seminal work of Ray $\&$ Vohra (1999) goes one step further, as the
worth of a coalition is deduced via arguments that satisfy a consistency criterion: a
deviant coalition takes into account the fact that after its deviation, other deviations might follow, with the newly deviant coalitions
thinking in a similar forward way. For games where binding agreements are feasible,
Huang $\&$ Sjostrom (1998, 2003) and Koczy (2007) developed the
recursive core. The recursive core is
constructed under the assumption that
the members of a coalition compute their value by looking recursively on the cores of the
sub-games played among the outsiders.

Predicting the equilibrium coalitional formation in a game with many players is computationally cumbersome.
Sandholm et.al (1999) showed that for an $n$-player game the number of different coalition structures
is $O(n^n)$ and $\omega(n^{\frac{n}{2}})$. Hence,
deducing the coalition structure that the outsiders form is a particularly difficult task (at least, for games
with a large number of players). For example, finding the coalition structure that maximizes the sum
of all players' payoffs is an $NP$-hard problem (Sandholm et.al 1999). Even finding sub-optimal solutions requires
the search of an exponential number of cases.

The last considerations give the motivation of the current paper. We analyze an $n$-firm
cooperative Cournot oligopoly assuming that no group of firms has the cognitive ability to accurately deduce the
partition that the outsiders will form. As a result, the members of a coalition cannot
compute their value with precision. Instead, they
compute it by following simple procedures or heuristics.

Clearly, the number of different heuristics
one can adopt is very large. Computer scientists, for example, model
similar situations via search algorithms that
give solutions within certain bounds from the optimal coalition structure
(Sandholm et.al 1999, Dang $\&$ Jennings 2004). On the other hand,
the economists' toolbox of heuristics includes models with players of various degrees of cognitive
abilities (Stahl $\&$ Wilson 1994, Camerer 2003, Camerer et.al 2004, Haruvy $\&$ Stahl 2007),
models with probabilistic choice
rules (McKelvey $\&$ Palfrey 1995, Chen et.al 1997, Anderson et.al 2002), to name only a few.

In our paper, the heuristics are based on the assignment of non-equilibrium probability
distributions over the set of the opponents' coalition structures.
I.e, when contemplating a deviation from
the grand coalition, the members of a coalition make the simplifying assumption
that the reactions of the outsiders are governed by various plausible -but not necessarily optimal-
probability distributions.

Our benchmark case assumes that the probability of a coalition structure
is  proportional to the profitability that
the structure induces for the outsiders. Namely, a deviant coalition assumes
it is more likely that its opponents will partition themselves according to
the more efficient structures. This approach is motivated by the logit
quantal response model of McKelvey $\&$ Palfrey (1995) in non-cooperative games, where the
probability of a player choosing a certain strategy depends on its relative payoff, with the probability
being positive even if the strategy is inferior.
We derive the characteristic function of a coalition under such a distribution -i.e, a logit type of distribution-
and we examine the core of the corresponding game.
We show that if the number of firms in the market is sufficiently large then the core
is non-empty. Hence, bounded rationality supports cooperation among all firms
in the market.

In the second part of the paper, we extend our analysis by considering more general
probability distributions. In particular we consider distributions which are related
via first-order stochastic dominance. If a certain distribution dominates another one, it gives relatively higher weight to partitions consisting of many coalitions. Our analysis has two goals: first, given that a relatively less concentrated partition hurts a deviant coalition, we present a novel way of modeling pessimism in a cooperative game with externalities. Secondly, we utilize this machinery to analyze the core in a Cournot market under a large number of distributions (other than the logit).

We fix a pair of distributions satisfying the first-order dominance property. We show that the core of the game under the dominated distribution is contained in the core of the game under the dominant. In particular, this implies that
the core under the logit distribution
is contained in the core under
any distribution that first-order dominates it. Thus we indirectly show that our Cournot game has a non-empty core for a large number of
probability distributions.

In particular, the above inclusion holds for the case of $\gamma$-core. Namely, the core under
the logit distribution is contained in the core constructed under the assumption
that outsiders form singleton coalitions (in our terminology, the $\gamma$ scenario corresponds to the degenerate distribution
that assigns probability one to the singletons partition). Hence our core refines the $\gamma$-core.

In what follows, we present the basic model in section 2. In sections 3 and 4 we present our results.
Section 5 concludes.

\section{The model}
We consider a market with the set $N=\{1,2,...,n\}$ of firms. Firms produce a homogeneous product. Firm $l$ produces quantity $q_l$ using the cost function $C(q_l)=cq_l,$
$l\in N.$ The market price $p$ is determined via the inverse demand function
$p=p(Q)$, where $Q=q_1+q_2+...+q_n$ is the total market quantity.

\vsp\noi\emph{Assumptions}

\vspace{0.1cm}\noi \emph{A1. $\exists$ $Q_0>0$ such that $p(Q)>0$ for $Q<Q_0$ and $p(Q)=0$ for $Q\geq Q_0$}

\vspace{0.2cm}\noi\emph{A2. $p'(Q)<0$, whenever $p(Q)>0$}

\vspace{0.2cm}\noi \emph{A3. $p'(Q)+q_ip''(Q)<0$}

\vsp\noi where $p'(Q)$ and $p''(Q)$ denote the first and second derivatives of the inverse demand function. The above assumptions are standard and guarantee the existence and uniqueness of Cournot equilibrium (see, for example, Vives 2001).

Let $S\subset N$ denote a coalition of firms with $|S|=s$ members and let $N\setminus S$ denote the complementary set of $S$, where $|N\setminus S|=n-s.$
The worth or value of $S$ is the sum of its members' profits. These profits depend on
how the members of $N\setminus S$ partition themselves into coalitions. The set $N\setminus S$ can be partitioned into
disjoint subsets in $B_{n-s}$ ways, where $B_{n-s}$ is Bell's $(n-s)^{th}$
number (Bell 1934).

What matters for $S$ is only the number of
the opponent coalitions in the Cournot market. Consider for example the case $N=\{1,2,3,4,5\}$ and $S=\{1\}$. The set of outsiders is $N\setminus S=\{2,3,4,5\}.$
Consider the partitions $\{\{2,3\},\{4,5\}\}$ and
$\{\{2,3,4\},\{5\}\}$ of outsiders. These partitions are equivalent for $S$ (and so are all partitions with two coalitions) in the sense that both induce the same profit for $S$ (in both cases, $S$ would compete in a triopoly market).
More generally, all partitions with $j$ coalitions
induce the same profit for $S$, irrespective of how the outsiders
are grouped among the $j$ coalitions. We will call these partitions \emph{$j$-similar}, where $j=1,2,...,n-s$.

Denote the number of $j$-similar partitions by
$K_{n-s,j}$, where $K_{n-s,j}$ gives the number of ways to
partition a set of $n-s$ objects into $j$ groups, or else the Stirling
number of the second kind. Then
\be\label{Stirling second kind} K_{n-s,j}=\frac{1}{j!}\sum\limits_{i=0}^j (-1)^i {{j}\choose{i}}(j-i)^{n-s} \ee

\vsp\noi The basic assumption that underlies this paper is that the members of $S$ use
simple probabilistic models in order to predict the coalitional behavior of the non-members.
As a benchmark case, we assume that the probability
of a partition is proportional to the profitability that the partition induces for the outsiders. This approach is in line
with the spirit of the logit quantal response
model (McKelvey $\&$ Palfrey 1995) in non-cooperative games, where the
probability of choosing a strategy depends on its relative payoff, the probability
being positive even if the strategy is inferior.

Consider again a coalition $S$ with $s$ members and an outsiders' partition with $j$ coalitions. Let $\Pi_j$ denote the sum of the profits that the $j$ coalitions
earn under this partition (this sum is constant
over all $j$-similar partitions).
Define

\be\label{log}f_{n,s}(j)=\frac{e^{\Pi_j}K_{n-s,j}}{\sum\limits_{m=1}^{n-s}
e^{\Pi_m}K_{n-s,m}}\ee
\vsp\noi Notice that $f_{n,s}(j)\in(0,1)$ and $\sum\limits_{j=1}^{n-s}f_{n,s}(j)=1$. Then, $f_{n,s}(j)$
gives the total probability that $S$ assigns to
all $j$-similar structures. Note in (\ref{log}) that the profitability of a $j$-similar partition is adjusted by the corresponding
Stirling number. The results of the paper hold even if such an adjustment does not take place.

\subsection{An example} Let us illustrate the above by considering an example with five firms, $N=\{1,2,3,4,5\}$. Assume that the inverse demand is $p=1-Q$ and that $c=0$. Consider again a coalition $S$ with $s$ members. If the $n-s$ outsiders form $j$ coalitions then there are $j+1$ active players in the market. By simple calculations, the total profits of the $j$ outside coalitions then are \be\label{sept}\Pi_j=\frac{j}{(j+2)^2},\hsp j=1,2,...,n-s\ee

\vspace{0.2cm}\noi Consider a singleton coalition, say $S=\{1\}.$ Then $B_{n-s}=B_4=15$ and $K_{4,1}=K_{4,4}=1, K_{4,2}=7, K_{4,3}=6$. Using (\ref{log}) and (\ref{sept}) the probabilities that $S$ assigns to outsiders' partitions are $$f_{5,1}(1)=f_{5,1}(4)=\frac{e^{1/9}}{Z_1},\hspace{0.2cm} f_{5,1}(2)=
\frac{7e^{1/8}}{Z_1}, \hspace{0.2cm} f_{5,1}(3)=
\frac{6e^{3/25}}{Z_1}$$

\vspace{0.2cm}\noi where $Z_1=2e^{1/9}+7e^{1/8}+6e^{3/25}.$ Consider next a coalition with two members, say $S=\{1,2\}$. Then $B_{n-s}=B_3=5$ and $K_{3,1}=K_{3,3}=1, K_{3,2}=3$. We then have $$f_{5,2}(1)=\frac{e^{1/9}}{Z_2}, \hspace{0.2cm}f_{5,2}(2)=\frac{3e^{1/8}}{Z_2}, \hspace{0.2cm}f_{5,2}(3)=\frac{e^{3/25}}{Z_2}$$

\vspace{0.2cm}\noi where $Z_2=e^{1/9}+3e^{1/8}+e^{3/25}.$  Consider next a coalition with three members $S=\{1,2,3\}$. In this case, $B_{n-s}=B_2=2$ and $K_{2,1}=K_{2,2}=1$. Hence we have $$f_{5,3}(1)=\frac{e^{1/9}}{Z_3},\hspace{0.35cm}f_{5,3}(2)=\frac{e^{1/8}}{Z_3}$$

\vspace{0.35cm}\noi where $Z_3=e^{1/9}+e^{1/8}.$ Finally, if a deviant coalition has four members, it faces one outsider only and so there is no ambiguity.

\subsection{The game $(N,v^n)$}
In this section we compute the characteristic function of a deviant coalition.
We use the $j$-similarity and focus for each $j$ on one
representative of the $j$-similar partitions. Let $q_s$ denote the quantity
of the deviant coalition $S$; and let $q_i^j$ denote the
quantity of outside coalition $i$, $i=1,2,...,j$, under a partition with $j$ members. The objective function that $S$ faces is given by

\be\label{prof S}\pi_f(S)=\sum\limits_{j=1}^{n-s}f_{n,s}(j)\big(p(q_s+\sum\limits_{i=1}^j q_i^j)-c\big)q_s\ee

\vsp\noi The objective function of coalition $i$ is
$$\pi_i^j=\big(p(q_s+\sum\limits_{r=1, r\neq i}^j q_r^j+q_i^j)-c\big)q_i^j,
\hspace{0.2cm}i=1,2,...,j$$

\vsp\noi Hence the maximization problems to
solve for are

\be\label{omo1}\max\limits_{q_s}\pi_f(S)\ee
\noi and for $j=1,2,...,n-s$,
\be\label{omo2}\max\limits_{q_i^j}\pi_i^j,\hspace{0.2cm}i=1,2,...,j\ee

\vsp\noi Let $\tilde{q}_s(f)=$ arg$\max\limits_{q_s}\pi_f(S)$ denote the best reply function of $S$;
and let $\tilde{q}_i^j=$ arg$\max\limits_{q_i^j}\pi_i^j$ denote the best reply{\footnote{Clearly the best replies depend on the quantities of the opponents but for notational simplicity we drop writing them.}} of coalition $i$, $i=1,2,...,j$ where $j=1,2,...,n-s$. If we solve the system of equations defined by the best replies we end up with the reduced-form solution in quantities, which shall be denoted by $q_s(f)$ and $q_i^j(f).$ By straightforward calculations, the reduced-form solution is given implicitly by

\be\label{qsffinal} q_s(f)=\frac{\sum\limits_{j=1}^{n-s}f_{n,s}(j)p(q_s(f)+jq_i^j(f))-c}{-\sum\limits_{j=1}^{n-s}f_{n,s}(j)p'(q_s(f)+jq_i^j(f))}\ee

\noi and for $j=1,2,...,n-s,$
\be\label{qjffinal} q_i^j(f)=\frac{p(q_s(f)+jq_i^j(f))-c}{-p'(q_s(f)+jq_i^j(f))},\hspace{0.2cm}\noi i=1,2,...,j\ee

\vspace{0.15cm}\noi where we used the fact that $q_1^j(f)=q_2^j(f)=...=q_j^j(f)$.
Using (\ref{qsffinal}) and (\ref{qjffinal}) in (\ref{prof S}),
we obtain the characteristic function of $S$. We denote this function by $v^n(S)$ or $v^n(s)$. Letting $Q_j(f)=q_s(f)+jq_i^j(f)$ we have

\vsp\be\label{fff worth S} v^n(S)=\frac{\big(\sum\limits_{j=1}^{n-s}f_{n,s}(j)p(Q_j(f))-c\big)^2}{-\sum\limits_{j=1}^{n-s}f_{n,s}(j)p'(Q_j(f))}
\ee
\vspace{0.15cm}\noi Hence our game is the pair $(N,v^n)$ where $v^n$ is defined by $(\ref{fff worth S})$. The value of the grand coalition is denoted by $v^n(N)$ or $v^n(n)$. This value is
the monopoly profit (which is independent of $n$ but for notational uniformity we keep the superscript). Denote by $Q_M$ the monopoly output. Then it is easy to show that
\vsp\noi $$v^n(N)=\frac{\big(p(Q_M)-c\big)^2}{-p'(Q_M)}$$

\vsp\noi An allocation is a vector $(x_1,x_2,...,x_n)$ such that $\sum\limits_{i\in N}x_i=v^n(N).$
The core ${\cal{C}}_f$ of $(N,v^n)$ is the set
of all allocations that cannot be blocked by any coalition, given distribution $f_{n,s}$.
I.e., the core is the set

\vspace{0.2cm}
$${\cal{C}}_f=\big\{(x_1,...,x_n): \not\exists S \hspace{0.2cm} with\hspace{0.2cm} v^n(S)>\sum\limits_{i\in S} x_i\big\}$$

\vspace{0.15cm}\noi Given the above, in the next sections we examine the core for various demand and probability functions.

\section{Results}

\noi The first result in this section states the following useful property.

\vsp\noi {\textbf{Lemma 1}} \emph{For every positive integer $k,$ the equality $v^n(s)=v^{n+k}(s+k)$ holds.}

\vspace{0.2cm}\noi \textbf{Proof} Consider two markets: in the first market there are $n$ firms and the deviant coalition $S$ has $s$ members and in the second there are $n+k$ firms and $S$ has $s+k$ members. Consider a number $j$ of outside coalitions. Notice that in the first market, $j$ runs from 1 up to $n-s$; in the second market, $j$ runs from 1 up to $n+k-(s+k)=n-s$ again. The total profits of $j$ outside coalitions under the first market are equal to their total profits under the second market. This is due to the constant returns to scale assumption. In both cases, these profits are $\Pi_j$.  Hence
\begin{align}
f_{n,s}(j)&=\frac{e^{\Pi_j}K_{n-s,j}}{\sum\limits_{m=1}^{n-s}e^{\Pi_m}K_{n-s,m}}\nonumber\\
&=\frac{e^{\Pi_j}K_{n+k-(s+k),j}}{\sum\limits_{m=1}^{n+k-(s+k)}e^{\Pi_m}K_{n+k-(s+k),m}}\nonumber\\
&=f_{n+k,s+k}(j)\
\end{align}

\vsp\noi Moreover for each $j$, $q_s(f)$ and $q_i^j(f)$ are constant under the two markets (again due to constant returns to scale): namely, the quantity of $S$ when it has $s$ members and the market has $n$ firms is equal to its quantity when $S$ has $s+k$ members and the market has $n+k$ firms; the same holds for $q_i^j(f)$ and by consequence for $Q_j(f)$. Combining this fact with (10) and (\ref{fff worth S}) proves the result.  \qed

\vsp\noi The intuition behind Lemma 1 is clear. If coalition $S$ has $s+k$
members in a market with $n+k$ firms, it faces
$n+k-(s+k)=n-s$ outsiders. This equals the number of outsiders
that $S$ faces when it has $s$ members in a market with $n$
firms. Hence $S$ faces the same set of potential coalition structures.

\vsp\noi An almost immediate implication of Lemma 1 is the monotonicity of $v^n(s)$ in $s$.

\vsp\noi {\textbf{Lemma 2}} \emph{For every $n$, $v^n(s)$ is strictly increasing in $s$.}

\vspace{0.2cm}\noi \textbf{Proof} The proof will be based on induction. For the base case, $n=2$, we have to prove that $v^2(2)>v^2(1)>v^2(0)$. Recall that $v^2(2)$ is the monopoly profit and $v^2(1)$ is the profit of a firm when two firms are in the market. Under assumptions {\emph{A1-A3}} we have that the former profit is higher than the latter,{\footnote{See Amir and Lambson (2000).}} i.e.,
$v^2(2)>v^2(1)$. Moreover, $v^2(1)>0=v^2(0)$ and so we have the base case.

Assume for the induction hypothesis that in a game with $n$ players and for an arbitrary $s$, we have that $v^n(s)>v^n(s-1)$.
We will prove that $v^{n+1}(s)>v^{n+1}(s-1)$. By Lemma 1 and the induction hypothesis we have that $v^{n+1}(s)=v^n(s-1)>v^n(s-2)=v^{n+1}(s-1)$. Note also that $v^{n+1}(s+1)>v^{n+1}(s)$  (by Lemma 1) and that $v^{n+1}(1)>v^{n+1}(0)=0$. This completes the proof. \qed

\vspace{0.2cm}\vsp\noi Lemmas 1 and 2 hold under any demand function (that satisfies assumptions  {\emph{A1-A3}}). In what follows, we use these Lemmas to derive conditions for core non-emptiness under certain demand functions. In particular we will focus on the family of demand functions $$Q=1-p^b, \hspace{0.2cm}b>0 $$ which we borrow from Anderson $\&$ Engers (1992). Note that if $b> (<)$ $1$, demand is concave (convex); and if $b=1$, demand is linear.
In order to derive analytically the market solution for an arbitrary $b$ we need to set $c=0$. The relevant calculations appear in Lemma A1 in the Appendix.
The solution, i.e, the quantities and the characteristic function, is given by
\vsp\noi \be\label{linqsffinal} q_s(f)=\frac{\sum\limits_{j=1}^{n-s}f_{n,s}(j)\psi_j^{\frac{1}{b}}}
{\sum\limits_{j=1}^{n-s}f_{n,s}(j)\psi_j^{\frac{1}{b}}(1+j+1/b)}\ee

\vsp\be\label{linqjffinal} q_i^j(f)=b\psi_j\frac{\sum\limits_{j=1}^{n-s}f_{n,s}(j)\psi_j^{\frac{1}{b}}(j+1/b)}{\sum\limits_{j=1}^{n-s}f_{n,s}(j)\psi_j^{\frac{1}{b}}(1+j+1/b)},\hspace{0.2cm}\noi i=1,2,...,j \ee
\vsp
\be\label{linfff worth S} v^n(S)=\sum\limits_{j=1}^{n-s}f_{n,s}(j)\Bigg(\frac{\sum\limits_{j=1}^{n-s}f_{n,s}(j)\psi_j^{\frac{1}{b}}(j+1/b)}
{(bj+1)\sum\limits_{j=1}^{n-s}f_{n,s}(j)\psi_j^{\frac{1}{b}}(1+j+1/b)}\Bigg)^{\frac{1}{b}}q_s(f)
\ee
\vsp\noi where ${\ds{\psi_j=\frac{1}{bj+1}}}$. Finally it is easy to see that the value of the grand coalition is $$v^n(N)=\frac{b}{(1+b)^{\frac{1}{b}+1}}$$

\subsection{The case $b=1$} As a benchmark case, we first present a result for the linear demand, i.e., $b=1.$ Afterwards, we discuss the non-linear case.

\vsp\noi \textbf{Proposition 1} \emph{Assume the demand function is given by $Q=1-p$. If $n$ is sufficiently large then ${\cal{C}}_f\neq\emptyset$.}

\vspace{0.2cm}\noi \textbf{Proof} Since firms are identical, the core is non-empty
if and only if for all $s \leq n,$

\be\label{9-Rajan}\frac{v^n(n)}{n}\geq \frac{v^n(s)}{s}\ee

\vspace{0.2cm}\noi It is easy to verify that the above inequality does not hold for $3 \leq n \leq 11$.
So for these values of $n$ the core is empty.\footnote{For $3\leq n\leq 11$
it holds that $v^n(1)>\frac{v^n(n)}{n}$ (see Table \ref{t} in the Appendix). The
relevant calculations were made using the Maple program and they are available by the authors
upon request.} The inequality holds for $n=12$ (Table \ref{h} in the Appendix).
We will prove the rest of the proposition using induction on $n$, where $n\geq 12$.

~\

\noi {\it{Base}}: Table \ref{h} in the Appendix establishes the base case ($n=12$).

~\

\noi {\it{Induction hypothesis}}: For all $S:|S|=s \leq n$, ${\ds{\frac{v^n(n)}{n}\geq\frac{v^n(s)}{s}}}$.

~\

\noi {\it{Induction step}}: We will show that for all $S:|S|=s \leq n+1$, $$\frac{v^{n+1}(n+1)}{n+1} \geq \frac{v^{n+1}(s)}{s}$$

~\

\noi By Lemma 1 we have that $v^{n+1}(s)=v^{n+1}((s-1)+1)=v^n(s-1)$ and also $v^{n+1}(n+1)=v^n(n)$.
So we have to show that \be\label{inductionequation1} \frac{v^n(n)}{n+1} \geq \frac{v^n(s-1)}{s} \ee

\vspace{0.15cm}\noi From the Induction hypothesis we have

\vspace{0.2cm}$$\label{inductionequation2} v^n(n) \geq \frac{n}{s-1} v^n(s-1)$$

\noi and thus

\vspace{0.1cm} \be\label{i1} (s-1)v^n(n)\geq nv^n(s-1)\ee

\noi Using Lemma 2,

\vspace{0.1cm}\be\label{i2}v^n(n)>v^n(s-1)\ee

\noi Adding (\ref{i1}) and (\ref{i2}) we have $$ sv^n(n)> (n+1)v^n(s-1)$$

\noi which implies that (\ref{inductionequation1}) holds. So we have the proof
for $n+1$ and thus the proposition is proved. \qed

\vsp\vsp\noi The monopoly profit is independent of the number of firms $n$. On the other hand,
$v^n(s)$ decreases in $n$. As a result, for sufficiently large $n$ the difference $v^n(n)/n-v^n(s)/s$  becomes
positive for all $s$ and the core is non-empty.

\subsection{The case $b\neq 1$}
In this section we discuss the non-emptiness of the core for the non-linear demand case. To this end, we will utilize the previous results, i.e., Lemmas 1 and 2, and Proposition 1. Recall that Lemmas 1 and 2 hold for any demand function. Furthermore, among the three steps of the induction proof of Proposition 1, i.e., base step, induction hypothesis and induction step, essentially only the base step depends on the demand function used. The other two steps work independently of the demand function (given of course the validity of the base step). Hence when extending Proposition 1 to cases where $b\neq 1$ we only need to ensure the validity of the base step. Namely, for a certain value of $b$ we need to find a specific number of firms that provides the base step of the induction argument (we refer the reader to page 9, in the proof of Proposition 1).

Due to the complexity of the model, we do not address the above task for all values of $b$. Nonetheless, Table \ref{nb} (next page) presents pairs of numbers $(b^*,n(b^*))$ that satisfy the property discussed above: given a certain value $b^*$ (where $b^*\neq 1$), the table reports a value $n(b^*)$ which is the number of firms that establishes the base step of the induction process for the demand function $Q=1-p^{b^*}$. In other words, given a specific value $b^*$, the game has non-empty core for $n\geq n(b^*)$ and it has empty core for $n<n(b^*)$.  We note that Table 1 provides values $b^*$ for which the demand function can be either convex ($b^*<1$) or concave ($b^*>1$).

\begin{table}[h!]
\begin{center}
\small

\begin{tabular}{|r|r|}
\hline
$b^*$&$n(b^*)$\\
\hline
0.5&3\\
0.6&5\\
0.7&6\\
0.8&7\\
0.9&9\\
1.1&15\\
1.2&19\\
1.3&24\\
1.4&32\\
1.5&42\\
1.6&57\\
1.7&78\\
1.8&107\\
1.9&147\\
2.0&205\\

\hline
\end{tabular}
\caption{$b^*$ and $n(b^*)$ for base step of induction.}
\label{nb}
\end{center}
\end{table}

In the Table we note that as $b^*$ increases, the core is non-empty less often (as $b^*$ increases, the number $n(b^*)$ increases). Moreover, if $b^*$ is sufficiently low, the core is non-empty for all $n\geq 3.$

\section{First-order stochastic dominance}
In this section we compare the cores of games that differ with respect to the probability
schemes assigned to outsiders' partitions. In particular, we consider distributions that are related via
first-order stochastic dominance. We will show that if a certain distribution dominates at first-order another one, then the core under the dominated distribution is a subset of the core under the dominant distribution. An application of this result is that Proposition 1 holds under any distribution that dominates the distribution defined in (\ref{log}).

Consider coalition $S$. Let $z_{n,s}$ and $w_{n,s}$ be two probability distributions over the set of the outsiders' partitions. Assume that $z_{n,s}$ dominates $w_{n,s}$ at first-order, i.e., for all $j^*,$

$$\sum\limits_{j=1}^{j^*}z_{n,s}(j)\leq\sum\limits_{j=1}^{j^*}w_{n,s}(j)$$

\vsp\noi Denote by ${\cal{C}}_w$ and ${\cal{C}}_z$ the cores under the two distributions and assume that ${\cal{C}}_z\neq\emptyset.$ We have the following result.

\vsp\noi{\textbf{Proposition 2}} \emph{Assume the inverse demand $p(Q)$ is weakly concave. If $z_{n,s}$ stochastically dominates $w_{n,s}$ at first-order
then ${\cal{C}}_w\subseteq {\cal{C}}_z$.}

\vspace{0.2cm}\noi {\textbf{Proof}} For $j=1,2,...,n-s$, denote $\tilde{Q}_j^{-s}=\sum\limits_{i=1}^j\tilde{q}_i^j$, where $\tilde{q}_i^j=\arg\max\limits_{q_i^j}\pi_i^j$ (recall that $\pi_i^j$ is the objective function of coalition $i$ in a partition with $j$ coalitions).
Let $\pi_w(S)$ and $\pi_z(S)$ denote the objective functions of $S$ under distributions $w_{n,s}$ and $z_{n,s}$ respectively. Let 

\vspace{0.1cm}$$\tilde{q}_s(w)=\arg\max\limits_{q_s}\pi_w(S),\hsp \tilde{q}_s(z)=\arg\max\limits_{q_s}\pi_z(S)$$

\vspace{0.1cm}\noi Denote by $q_s(w)$, $q_i^j(w)$ and $Q_{j}^{-s}(w)=\sum\limits_{i=1}^jq_i^j(w)$ the reduced-form solution of the system of equations defined by the above best replies
when the probability distribution is $w_{n,s}$; and by $q_s(z)$, $q_i^j(z)$ and $Q_{j}^{-s}(z)=\sum\limits_{i=1}^jq_i^j(z)$ the reduced-form solution when the probability distribution is $z_{n,s}$. Finally, denote by
$v_w^n(S)$ and $v_z^n(S)$ the characteristic functions of coalition $S$ under $w_{n,s}$ and $z_{n,s}$ respectively.

We first have that
\vspace{0.25cm} $$\tilde{Q}_{j+1}^{-s}>\tilde{Q}_{j}^{-s},\hspace{0.2cm} j=1,2,...,n-s-1$$

\vspace{0.25cm}\noi This follows by Amir $\&$ Lambson (2000, Theorem 2.2 (b)). Next define
$$\tilde{\pi}_w(S)=\sum\limits_{j=1}^{n-s}w_{n,s}(j)\big(p(q_s+\tilde{Q}_j^{-s})-c\big)q_s\equiv
\sum\limits_{j=1}^{n-s}w_{n,s}(j)\pi_s(q_s,\tilde{Q}_j^{-s})$$ and $$\tilde{\pi}_z(S)=\sum\limits_{j=1}^{n-s}z_{n,s}(j)\big(p(q_s+\tilde{Q}_j^{-s})-c\big)q_s
\equiv\sum\limits_{j=1}^{n-s}z_{n,s}(j)\pi_s(q_s,\tilde{Q}_j^{-s})$$

\vsp\noi where $$\pi_s(q_s,\tilde{Q}_j^{-s})\equiv \big(p(q_s+\tilde{Q}_j^{-s})-c\big)q_s,\hspace{0.25cm} j=1,2,...,n-s$$

\vsp\noi Notice that since $\tilde{Q}_{j+1}^{-s}>\tilde{Q}_{j}^{-s}$, we have \be\label{im} \pi_s(q_s,\tilde{Q}_j^{-s})>\pi_s(q_s,\tilde{Q}_{j+1}^{-s})
,\hspace{0.2cm} j=1,2,...,n-s-1\ee

\vsp\noi where the last inequality is due to the fact that the profit of a firm in a Cournot market is decreasing in the (individual or aggregate) quantities of the other firms. Notice next that

\begin{align}
\tilde{\pi}_w(S)-\tilde{\pi}_z(S)&= \big(\underbrace{w_{n,s}(1)-z_{n,s}(1)}_{\geq 0}\big)\pi_s(q_s,\tilde{Q}_1^{-s})+\sum\limits_{j=2}^{n-s}
\big(w_{n,s}(j)-z_{n,s}(j)\big)\pi_s(q_s,\tilde{Q}_j^{-s})\nonumber\\
&> \big(w_{n,s}(1)-z_{n,s}(1)\big)\pi_s(q_s,\tilde{Q}_2^{-s})+\sum\limits_{j=2}^{n-s}
\big(w_{n,s}(j)-z_{n,s}(j)\big)\pi_s(q_s,\tilde{Q}_j^{-s})\nonumber\\
&=\big(\underbrace{w_{n,s}(1)+w_{n,s}(2)-z_{n,s}(1)-z_{n,s}(2)}_{\geq 0}\big)\pi_s(q_s,\tilde{Q}_2^{-s})\nonumber\\
&+\sum\limits_{j=3}^{n-s}
\big(w_{n,s}(j)-z_{n,s}(j)\big)\pi_s(q_s,\tilde{Q}_j^{-s}))\
\end{align}

\vsp\noi where the inequality is due to (\ref{im}). Continuing the iterations on $j$, we eventually have that

\begin{align}
\tilde{\pi}_w(S)-\tilde{\pi}_z(S)&>\big(\underbrace{\sum\limits_{j=1}^{n-s-1}w_{n,s}(j)-\sum\limits_{j=1}^{n-s-1}z_{n,s}(j)}_{\geq 0}\big)\pi_s(q_s,\tilde{Q}_{n-s-1}^{-s}) \nonumber\\
&+\big(w_{n,s}(n-s)-z_{n,s}(n-s)\big)\pi_s(q_s,\tilde{Q}_{n-s}^{-s})\nonumber\\
&>\big(\underbrace{\sum\limits_{j=1}^{n-s}w_{n,s}(j)-\sum\limits_{j=1}^{n-s}z_{n,s}(j)}_{= 0}\big)\pi_s(q_s,\tilde{Q}_{n-s}^{-s})=0\
\end{align}

\vsp\noi where we again used (\ref{im}). We conclude that \emph{for all} $q_s$ and the \emph{corresponding} $\tilde{Q}_j^{-s}$, we have \be\label{sos}\sum\limits_{j=1}^{n-s}
w_{n,s}(j)\pi_s(q_s,\tilde{Q}_j^{-s})>\sum\limits_{j=1}^{n-s}
z_{n,s}(j)\pi_s(q_s,\tilde{Q}_j^{-s})\ee

\vsp\noi In the Appendix we show that if the inverse demand function $p(Q)$ is weakly concave then $Q_{j}^{-s}(w)< Q_{j}^{-s}(z)$ (Lemma A2 in the Appendix).
Notice next that

\begin{align}
v_w^n(S)&=\sum\limits_{j=1}^{n-s}
w_{n,s}(j)\pi_s(q_s(w),Q_j^{-s}(w))\nonumber\\
&\geq \sum\limits_{j=1}^{n-s}
w_{n,s}(j)\pi_s(q_s(z),Q_j^{-s}(w))\nonumber\\
&\geq \sum\limits_{j=1}^{n-s}
w_{n,s}(j)\pi_s(q_s(z),Q_j^{-s}(z))\nonumber\\
&>\sum\limits_{j=1}^{n-s}
z_{n,s}(j)\pi_s(q_s(z),Q_j^{-s}(z))=v_z^n(S)\
\end{align}

\vsp\noi where the first inequality is due to the fact that $q_s(w)$ is the optimal choice of coalition $S$ against $Q_j^{-s}(w), j=1,2,...,n-s$; the second inequality holds because $Q_j^{-s}(w)< Q_j^{-s}(z)$ and because the Cournot profit of a firm decreases in the quantities of the other firms; and the third inequality is due to (\ref{sos}). Since $v_w^n(S)>v_z^n(S)$ we conclude that ${\cal{C}}_w\subseteq {\cal{C}}_z$.
\qed

\vsp\noi As an application of Proposition 2, we note that Proposition 1 holds not only under $f_{n,s}$ but also under any distribution that dominates $f_{n,s}$ at first order.

\vsp\noi {\textbf{Corollary 1}} \emph{Assume the demand function is given by $Q=1-p$. Consider any distribution $z_{n,s}$ that dominates distribution (\ref{log}) at first-order. Then ${\cal{C}}_z\neq\emptyset$ if $n$ is sufficiently large.}

\vspace{0.2cm}\noi {\textbf{Proof}} If demand is linear and $n$ is large then ${\cal{C}}_f\neq\emptyset$ (Proposition 1). Since $z_{n,s}$ dominates distribution (\ref{log}) at first-order then ${\cal{C}}_f\subseteq {\cal{C}}_z$. Hence the latter core is non-empty.\qed

\vsp\noi Compared to the cumulative distribution of $f_{n,s}$, the cumulative distribution of $z_{n,s}$
assigns higher probabilities to events that include partitions
with many coalitions. Clearly, these events are unfavorable for $S$, hence the use of $z_{n,s}$ indicates some sort of pessimism (relatively to $f_{n,s})$  on behalf of the members of $S$. This approach can be motivated by resorting to
the theory of risk measurement, which often measures risk by assigning relatively high probabilities to unfavorable events (see e.g, Acerbi 2002).\footnote{We thank an anonymous referee for pointing out this connection.} Our analysis provides a novel way of tracing the impact of pessimism or risk in a cooperative game with externalities.

Finally we note that a particular distribution that dominates $f_{n,s}$ is the distribution defined
by $\wh{z}_{n,s}(j)=0$, for $j=1,2,...,n-s-1$ and $\wh{z}_{n,s}(n-s)=1.$
This distribution corresponds to the $\gamma$-core
scenario. It is known that the latter core is non-empty for general Cournot oligopolies (Chander 2010).
Letting ${\cal{C}}_{\wh{z}}$ denote the $\gamma$-core, we have the following corollary.

\vsp\noi {\textbf{Corollary 2}} \emph{The inclusion ${\cal{C}}_f\subseteq {\cal{C}}_{\wh{z}}$ holds.}

\vsp\noi The $\gamma$-core is based on the worst scenario for
$S$: all $n-s$ firms remain separate
entities. Under $f_{n,s}$, the singleton coalitions structure is just one
of the partitions that $S$ takes into account. Other, more favorable, partitions occur with positive probability. Hence,
the value of $S$ under $f_{n,s}$ is higher than its value under $\wh{z}_{n,s}$.

Corollaries 1 and 2 are stated in terms of the linear demand function.
We note that similar statements hold for the non-linear demand case (we refer the reader to section 3.2 that presents cases of non-linear demand functions for which the core is non-empty).

\section{Conclusions}
This paper analyzed cooperative Cournot games. The analysis is based on the assertion
that when a coalition contemplates a deviation from the grand coalition, it assigns various non-equilibrium
distributions on the set of partitions that the outsiders can form. This assumption is justified by
imposing cognitive constraints on behalf of the firms in the market.
Provided that the number of firms is sufficiently high, the core is non-empty for a large number of probability distributions and demand functions.

Let us mention a few extensions of the current work.
The analysis of oligopolistic markets with more general cost functions
and/or other modes of competition (e.g., product differentiation, price competition)
are natural future directions.
Furthermore, the application of
the current framework to other economic environments or to abstract cooperative games with externalities
is of interest.

\newpage
\section*{Appendix}

\begin{table}[h]
\begin{center}
\small

\begin{tabular}{|r|r|r|}
\hline
$s$&$v^n(s)$&$v^n(s)/s$\\
\hline
1&0.02047&0.02047\\
2&0.02273&0.01136\\
3&0.02544&0.00848\\
4&0.02876&0.00719\\
5&0.03289&0.00657\\
6&0.03815&0.00635\\
7&0.04503&0.00643\\
8&0.05443&0.00680\\
9&0.06795&0.00755\\
10&0.08736&0.00873\\
11&0.11111&0.01010\\
12&0.25000&0.02083\\

\hline
\end{tabular}
\caption{values $v^n(s)$ and $v^n(s)/s$ with $n=12$}
\label{h}
\end{center}
\end{table}

\begin{table}[t]
\begin{center}
\small

\begin{tabular}{|r|r|r|}
\hline
$n$&$v^n(1)$&$v^n(n)/n$\\
\hline
3&0.08736&0.08333\\
4&0.06795&0.06250\\
5&0.05444&0.05000\\
6&0.04503&0.04166\\
7&0.03815&0.03571\\
8&0.03289&0.03125\\
9&0.02876&0.02777\\
10&0.02544&0.02500\\
11&0.02273&0.02272\\

\hline
\end{tabular}
\caption{values $v^n(1)$ and $v^n(n)/n,$ $n\in\{3,4,...,11\}$}
\label{t}
\end{center}
\end{table}

\newpage

\vsp\noi {\textbf{Lemma A1}}\emph{ Assume the demand function is $Q=1-p^b$. Then the characteristic function is given by (\ref{linfff worth S}).}

\vspace{0.2cm}\noi {\textbf{Proof}} The profit function of coalition $i$ under a partition with $j$ members is
\be\label{non2}\pi_i^j=(1-q_s-\sum\limits_{r=1, r\neq i}^j q_r^j-q_i^j)^{\frac{1}{b}}q_i^j,
\hspace{0.3cm}i=1,2,...,j\ee

\vsp\noi Note that \be\label{non3}\frac{\partial{\pi_i^j}}{\partial{q_i^j}}=0\Leftrightarrow b(1-q_s-q_i^j-\sum\limits_{r=1, r\neq i}^j q_r^j)=q_i^j\ee
By symmetry, all $j$ outside coalitions produce the same. So let $q_r^j=q_i^j$, for all $r$. Therefore by (\ref{non3}), $b(1-q_s-jq_i^j)=q_i^j$ and hence \be\label{non4}\tilde{q}_i^j=\frac{b(1-q_s)}{bj+1}\ee

\vspace{0.2cm}\noi The objective function of the deviant coalition $S$ is \be\label{nonminus1}\pi_f(S)= \sum\limits_{j=1}^{n-s}f_{n,s}(j)
(1-q_s-\sum\limits_{i=1}^j q_i^j)^{\frac{1}{b}}q_s \ee

\vspace{0.2cm}\noi Note that ${\ds{\frac{\partial{\pi_f(S)}}{\partial{q_s}}=0}}$ if

\be\label{non5}\sum\limits_{j=1}^{n-s}f_{n,s}(j)(1-q_s-\sum_{i=1}^jq_i^j)^{\frac{1}{b}}=
\frac{1}{b}\sum\limits_{j=1}^{n-s}f_{n,s}(j)(1-q_s-\sum_{i=1}^jq_i^j)^{\frac{1}{b}-1}q_s\ee
\vsp\noi Using (\ref{non4}), (\ref{non5}) becomes

$$\sum\limits_{j=1}^{n-s}f_{n,s}(j)\big(\frac{1-q_s}{1+bj}\big)^{\frac{1}{b}}=
\frac{1}{b}\sum\limits_{j=1}^{n-s}f_{n,s}(j)\big(\frac{1-q_s}{1+bj}\big)^{\frac{1}{b}-1}q_s$$

\vsp\noi and hence
$$(1-q_s)^{\frac{1}{b}}\sum\limits_{j=1}^{n-s}f_{n,s}(j)\big(\frac{1}{1+bj}\big)^{\frac{1}{b}}=
\frac{1}{b}(1-q_s)^{\frac{1}{b}-1}\sum\limits_{j=1}^{n-s}f_{n,s}(j)\big(\frac{1}{1+bj}\big)^{\frac{1}{b}-1}q_s$$

\vsp\noi Define ${\ds{\psi_j=\frac{1}{bj+1}.}}$ Then rearranging the above relation gives \be\label{au0}q_s\big(\sum\limits_{j=1}^{n-s}f_{n,s}(j)\psi_j^{\frac{1}{b}}
+\frac{1}{b}\sum\limits_{j=1}^{n-s}f_{n,s}(j)\psi_j^{\frac{1}{b}-1}\big)=\sum\limits_{j=1}^{n-s}f_{n,s}(j)\psi_j^{\frac{1}{b}}\ee

\vsp\noi Notice that $$\sum\limits_{j=1}^{n-s}f_{n,s}(j)\psi_j^{\frac{1}{b}}
+\frac{1}{b}\sum\limits_{j=1}^{n-s}f_{n,s}(j)\psi_j^{\frac{1}{b}-1}=\sum\limits_{j=1}^{n-s}f_{n,s}(j)\psi_j^{\frac{1}{b}}(1+\frac{1}{b\psi_j})=$$

\be\label{au1}\sum\limits_{j=1}^{n-s}f_{n,s}(j)\psi_j^{\frac{1}{b}}(1+(bj+1)/b)=\sum\limits_{j=1}^{n-s}f_{n,s}(j)\psi_j^{\frac{1}{b}}(1+j+1/b)\ee

\vsp\noi Using (\ref{au0}) and (\ref{au1}) we get \be\label{non9} q_s(f)=\frac{\sum\limits_{j=1}^{n-s}f_{n,s}(j)\psi_j^{\frac{1}{b}}}
{\sum\limits_{j=1}^{n-s}f_{n,s}(j)\psi_j^{\frac{1}{b}}(1+j+1/b)} \ee

\vsp\noi  Using (\ref{non9}), (\ref{non4}) becomes

\be\label{nnnn} q_i^j(f)=b\psi_j\frac{\sum\limits_{j=1}^{n-s}\psi_j^
{\frac{1}{b}}(j+1/b)}{\sum\limits_{j=1}^{n-s}f_{n,s}(j)\psi_j^{\frac{1}{b}}(1+j+1/b)}\ee

\vsp\noi Plugging (\ref{non9}) and (\ref{nnnn}) in (\ref{nonminus1}) gives us (\ref{linfff worth S}).\qed

\vsp\noi {\textbf{Lemma A2}} \emph{Assume the inverse demand $p(Q)$ is weakly concave. Then $Q_{j}^{-s}(w)< Q_j^{-s}(z)$, $j=1,2,...,n-s$.}

\vspace{0.2cm}\noi {\textbf{Proof}} Fix $Q_j^{-s}$. For notational convenience, define $F(q_s)\equiv\sum\limits_{j=1}^{n-s}w_{n,s}(j)\pi_s(q_s,Q_j^{-s})$ and $H(q_s)\equiv\sum\limits_{j=1}^{n-s}z_{n,s}(j)\pi_s(q_s,Q_j^{-s})$. Then $\tilde{q}_s(w)$ and $\tilde{q}_s(z)$ satisfy
respectively the first-order conditions
${\ds{\frac{\partial{F(q_s)}}{\partial{q_s}}=0}}$ and ${\ds{\frac{\partial{H(q_s)}}{\partial{q_s}}=0}}$ or equivalently

\vsp\be\label{ub01}\sum\limits_{j=1}^{n-s}w_{n,s}(j)p'(q_s+Q_{j}^{-s})q_s+\sum\limits_{j=1}^{n-s}w_{n,s}(j)p(q_s+Q_{j}^{-s})-c=0\ee and

\be\label{ub02}\sum\limits_{j=1}^{n-s}z_{n,s}(j)p'(q_s+Q_{j}^{-s})q_s+\sum\limits_{j=1}^{n-s}z_{n,s}(j)p(q_s+Q_{j}^{-s})-c=0\ee

\vsp\noi The function $F(q_s)$ is strictly concave in $q_s$ (by assumptions A1-A3). Hence $\tilde{q}_s(w)>\tilde{q}_s(z)$ if and only if ${\ds{\frac{\partial{F(\tilde{q}_s(z))}}{\partial{q_s}}>0}}$. By (\ref{ub01}) we have that

\be\label{ub03}\frac{\partial{F(\tilde{q}_s(z))}}{\partial{q_s}}>0\Leftrightarrow \sum\limits_{j=1}^{n-s}w_{n,s}(j)p'(\tilde{q}_s(z)+Q_{j}^{-s})\tilde{q}_s(z)+\sum\limits_{j=1}^{n-s}w_{n,s}(j)p(\tilde{q}_s(z)+Q_{j}^{-s})-c>0\ee

\vsp\noi Solving for $\tilde{q}_s(z)$ by (\ref{ub02}) and plugging in (\ref{ub03}) we have that

\be\label{ub04}\frac{\partial{F(\tilde{q}_s(z))}}{\partial{q_s}}>0\Leftrightarrow \frac{\sum\limits_{j=1}^{n-s}w_{n,s}(j)p'(\tilde{q}_s(z)+Q_{j}^{-s})}{-\sum\limits_{j=1}^{n-s}z_{n,s}(j)p'(\tilde{q}_s(z)+Q_{j}^{-s})}
\big(\sum\limits_{j=1}^{n-s}z_{n,s}(j)p(\tilde{q}_s(z)+Q_{j}^{-s})-c\big)$$

$$+\sum\limits_{j=1}^{n-s}w_{n,s}(j)p(\tilde{q}_s(z)+Q_{j}^{-s})-c>0\ee

\vsp\noi We now use the concavity assumption and claim that \be\label{ub05}\frac{\sum\limits_{j=1}^{n-s}w_{n,s}(j)p'(\tilde{q}_s(z)+Q_{j}^{-s})}{-\sum\limits_{j=1}^{n-s}z_{n,s}(j)
p'(\tilde{q}_s(z)+Q_{j}^{-s})}\geq -1\ee

\vsp\noi To show the above we can equivalently show \be\label{ub06}\sum\limits_{j=1}^{n-s}w_{n,s}(j)p'(\tilde{q}_s(z)+Q_{j}^{-s})-
\sum\limits_{j=1}^{n-s}z_{n,s}(j)p'(\tilde{q}_s(z)+Q_{j}^{-s})\geq 0\ee

\vsp\noi We have $$\sum\limits_{j=1}^{n-s}w_{n,s}(j)p'(\tilde{q}_s(z)+Q_{j}^{-s})-
\sum\limits_{j=1}^{n-s}z_{n,s}(j)p'(\tilde{q}_s(z)+Q_{j}^{-s})=$$

$$\big(w_{n,s}(1)-z_{n,s}(1)\big)p'(\tilde{q}_s(z)+Q_1^{-s})+\sum\limits_{j=2}^{n-s}\big(w_{n,s}(j)-
z_{n,s}(j)\big)p'(\tilde{q}_s(z)+Q_{j}^{-s})\geq $$

$$\big(w_{n,s}(1)-z_{n,s}(1)\big)p'(\tilde{q}_s(z)+Q_2^{-s})+\sum\limits_{j=2}^{n-s}\big(w_{n,s}(j)-
z_{n,s}(j)\big)p'(\tilde{q}_s(z)+Q_{j}^{-s})$$

\vsp\noi where the inequality holds because $w_{n,s}(1)-z_{n,s}(1)\geq 0$ and because the weak concavity of price implies that $p'(\tilde{q}_s(z)+Q_{1}^{-s})\geq p'(\tilde{q}_s(z)+Q_{2}^{-s})$ (recall that $Q_1^{-s}<Q_{2}^{-s}).$ If we continue the process of iterating $j$, we end up with (\ref{ub06}). Since the latter condition holds, we have that

$$\frac{\sum\limits_{j=1}^{n-s}w_{n,s}(j)p'(\tilde{q}_s(z)+Q_{j}^{-s})}{-\sum\limits_{j=1}^{n-s}z_{n,s}(j)p'(\tilde{q}_s(z)+Q_{j}^{-s})}
\big(\sum\limits_{j=1}^{n-s}z_{n,s}(j)p(\tilde{q}_s(z)+Q_{j}^{-s})-c\big)+\sum\limits_{j=1}^{n-s}w_{n,s}(j)p(\tilde{q}_s(z)+Q_{j}^{-s})-c\geq $$

$$-\big(\sum\limits_{j=1}^{n-s}z_{n,s}(j)p(\tilde{q}_s(z)+Q_{j}^{-s})-c\big)+\sum\limits_{j=1}^{n-s}w_{n,s}(j)p(\tilde{q}_s(z)+Q_{j}^{-s})-c=$$

\be\label{ub07}\sum\limits_{j=1}^{n-s}w_{n,s}(j)p(\tilde{q}_s(z)+Q_{j}^{-s})-\sum\limits_{j=1}^{n-s}z_{n,s}(j)p(\tilde{q}_s(z)+Q_{j}^{-s})\ee

\vsp\noi But expression (\ref{ub07}) can be written as
$$\big(w_{n,s}(1)-z_{n,s}(1)\big)p(\tilde{q}_s(z)+Q_1^{-s})+\sum\limits_{j=2}^{n-s}\big(w_{n,s}(j)-z_{n,s}(j)\big)p(\tilde{q}_s(z)+Q_{j}^{-s})>$$

$$\big(w_{n,s}(1)-z_{n,s}(1)\big)p(\tilde{q}_s(z)+Q_2^{-s})+\sum\limits_{j=2}^{n-s}\big(w_{n,s}(j)-z_{n,s}(j)\big)p(\tilde{q}_s(z)+Q_{j}^{-s})$$

\vsp\noi where the last inequality holds because $Q_1^{-s}<Q_2^{-s}$. Continuing the iterations on $j$, we end up with
\be\label{ub08}\sum\limits_{j=1}^{n-s}w_{n,s}(j)p(\tilde{q}_s(z)+Q_{j}^{-s})-\sum\limits_{j=1}^{n-s}z_{n,s}(j)p(\tilde{q}_s(z)+Q_{j}^{-s})>0\ee

\vsp\noi Combining (\ref{ub04}), (\ref{ub07}) and (\ref{ub08}) we conclude that ${\ds{\frac{\partial{F(\tilde{q}_s(z))}}{\partial{q_s}}>0}}$ and hence $\tilde{q}_s(w)>\tilde{q}_s(z).$ But then $Q_{j}^{-s}(w)<Q_j^{-s}(z)$, since $Q_j^{-s}(w)$ and $Q_j^{-s}(z)$ emerge from $\tilde{Q}_j^{-s}$ for $q_s=\tilde{q}_s(w)$ and $q_s=\tilde{q}_s(z)$ respectively and commodities in a Cournot market are substitutes.\qed

\section*{References}

\begin{enumerate}[1.]

\item Acerbi, C. (2002) Spectral measures of risk: A coherent representation of subjective
risk aversion, Journal of Banking $\&$ Finance, 26, 1505-1518.

\item Amir, R. and V.E. Lambson (2000) On the effects of entry in Cournot markets, Review of Economic Studies, 67, 235-254.

\item Anderson S. and M.P Engers (1992) Stackelberg versus Cournot oligopoly equilibrium, International Journal of Industrial Organization, 10, 127-135.

\item Anderson S., J. K. Goeree and C. A. Holt (2002)
The logit equilibrium: A perspective on intuitive behavioral anomalies, Southern Economic Journal,
69, 21-47.

\item Aumann R. (1959) Acceptable points in general cooperative n-person games,
Contributions to the theory of games IV, Annals of Mathematics
Studies vol 40, Princeton University Press, Princeton.

\item Bell E. T. (1934) Exponential Numbers.
American Mathematical Monthly, 41, 411-419.

\item Camerer C. (2003) Behavioral studies of strategic thinking in games, Trends in Cognitive
Sciences, 7, 225-231.

\item Camerer C., T-H Ho and J-K Chong (2004) A cognitive hierarchy model of games, Quarterly Journal
of Economics, 119, 861-898.

\item Chander P. (2010) Cores of games with positive externalities, CORE Discussion Paper.

\item Chander P. and H. Tulkens (1997) A core of an economy with multilateral environmental
externalities, International Journal of Game Theory, 26, 379-401.

\item Chen H.C, J. Friedman and J.F. Thisse (1997)
Boundedly rational Nash equilibrium: a probabilistic choice approach.
Games and Economic Behavior, 18, 32-54.

\item Currarini S. and M. Marini (2003) A sequential approach to the characteristic
function and the core in games with externalities, Sertel, M., Kara, A.(eds.), Advances in Economic Design. Springer Verlag, Berlin.

\item Dang V. D. and N.R. Jennings (2004)
Generating coalition structures with finite bound from the optimal guarantees. In: 3rd International Conference
on Autonomous Agents and Multi-Agent Systems, New York, USA, 564-571.

\item Hart S. and M. Kurz (1983) Endogenous formation of coalitions, Econometrica, 52,
1047-1064.

\item Haruvy E. and D. Stahl (2007) Equilibrium selection and bounded rationality in
symmetric normal form games, Journal of Economic Behavior and Organization, 62, 98-119.

\item Huang C.Y. and T. Sjostrom (1998) The $p$-core, Working Paper.

\item Huang C.Y. and T. Sjostrom (2003) Consistent solutions for cooperative games
with externalities, Games and Economic Behavior, 43, 196-213.

\item Koczy L. (2007) A recursive core for partition function form games, Theory and Decision,
63, 41-51.

\item McKelvey R. and T. R. Palfrey (1995) Quantal response equilibria for normal
form games, Games and Economic Behavior 10, 6-38.

\item Rajan R. (1989)
Endogenous coalition formation in cooperative oligopolies, International Economic Review, 30, 863-876.

\item Ray D. and R. Vohra (1999) A theory of endogenous coalition structures,
Games and Economic Behavior, 26, 286-336.

\item Sandholm T., K. Larson, M. Andersson, O. Shehory and F. Tohm\'{e} (1999) Coalition
structure generation with worst case guarantees. Artificial Intelligence
111, 209-238.

\item Stahl D. and P. Wilson (1994) On players' models of other players: Theory and experimental evidence,
Games and Economic Behavior, 10, 218-254.

\item Vives X., (2001) Oligopoly Pricing: Old ideas and new tools, MIT Press.

\item Zhao, J. (1999) A $\beta$-core existence result and its application to oligopoly markets, Games
and Economic Behavior, 27, 153-168.

\end{enumerate}

\end{document}